\documentclass[fleqn,usenatbib]{mnras}

\usepackage{newtxtext,newtxmath}

\usepackage[T1]{fontenc}
\usepackage{ae,aecompl}

\usepackage{graphicx}	
\usepackage{amsmath}	
\usepackage[export]{adjustbox}
\usepackage{caption}
\usepackage{subcaption}

\newcommand{\approxprop}{\mathrel{\vcenter{
  \offinterlineskip\halign{\hfil$##$\cr
    \propto\cr\noalign{\kern2pt}\sim\cr\noalign{\kern-2pt}}}}}

\title[Eccentric debris disc morphologies II: combining free and forced eccentricities]{Eccentric debris disc morphologies II: Surface brightness variations from overlapping orbits in narrow eccentric discs}

\author[J. B. Lovell et al.]{
Joshua B. Lovell$^{1}$ \thanks{E-mail: joshualovellastro@gmail.com} and
Elliot M. Lynch$^{2}$
\\
$^{1}$Center for Astrophysics, Harvard \& Smithsonian, 60 Garden Street, Cambridge, MA 02138-1516, USA \\
$^{2}$Univ Lyon, Univ Lyon1, Ens de Lyon, CNRS, Centre de Recherche Astrophysique de Lyon UMR5574, F-69230, Saint-Genis,-Laval, France \\
}

\date{Accepted 7 Jun 2023. Received 26 Mar 2023.}

\pubyear{2023}

\begin{document}
\label{firstpage}
\pagerange{\pageref{firstpage}--\pageref{lastpage}}
\maketitle

\begin{abstract}
We present Paper II of the Eccentric Debris Disc Morphologies series to explore the effects that significant free and forced eccentricities have on high-resolution millimetre-wavelength observations of debris discs, motivated by recent ALMA images of HD53143’s disc.
In this work, we explore the effects of free eccentricity, and by varying disc fractional widths and observational resolutions, show for a range of narrow eccentric discs, orbital overlaps result in dust emission distributions that have either one or two radial peaks at apocentre and/or pericentre.
The narrowest discs contain two radial peaks, whereas the broadest discs contain just one radial peak.
For fixed eccentricities, as fractional disc widths are increased, we show that these peaks merge first at apocentre (producing apocentre glow), and then at pericentre (producing pericentre glow).
Our work thus demonstrates that apocentre/pericentre glows in models with constant free and forced eccentricities can be both width and resolution dependent at millimetre wavelengths, challenging the classical assertion that apocentre/pericentre glows are purely wavelength dependent.
We discuss future high-resolution observations that can distinguish between competing interpretations of underlying debris disc eccentricity distributions.
\end{abstract}



\vspace{-0.2in}
\section{Introduction}
Debris discs are a relatively common type of circumstellar disc, with instruments such as Spitzer and Herschel detecting these towards about 20--30\% of main-sequence stars \citep{Wyatt08, Hughes18}. 
Their presence indicates that belts of planetesimals have formed and are evolving via collisional erosion, producing observable dust grains which trace the orbits of their parent bodies.

The observed surface brightness distributions of debris disc dust can be used to infer the distribution of planetesimals and planets within planetary systems.
For example, as planets sweep up and/or scatter nearby planetesimals, these can result in disc cavities, the structure of which can be used to constrain the type of planet/s responsible \citep{Shannon16}, important for understanding the population of planets on wide orbits in a parameter space far outside of current planet-detection capabilities.
In addition, bodies orbiting nearby to eccentric planets \citep[i.e., beyond the planetary clearing zone][]{Quillen06} can have their own orbits \textit{shaped} as a result of three-body dynamics between the star and planet, and individual planetesimals \citep[see e.g.,][]{Murray99, Wyatt99, Wyatt05}.
This effect has been interpreted as the origin of the eccentric debris disc structure observed in the discs of Fomalhaut, HR\,4796, HD202628, HD38206, and most recently HD53143, \citep[see][]{MacGregor13, Olofsson2019, Faramaz19, Booth21, MacGregor22}.

The classical paradigm in debris discs was developed in the context of low-resolution data which often captured disc ansae within single beams.
This picture found that eccentric debris disc emission structures could be largely understood as a function of wavelength.
\citet{Wyatt99} demonstrated that at mid-to-far-infrared wavelengths, eccentric debris discs exhibit `pericentre glow', whereby the warmest dust (closest to the star) produces enhanced emission relative to the coolest dust (furthest from the star). 
The predictions of this theory match Keck and Herschel observations of discs \citep[e.g., HR4796A and Fomalhaut; see][]{Telesco00, Acke12}.
Conversely, at longer submillimetre/millimetre wavelengths observations are more sensitive to the dust mass under the beam. By modelling 1-dimensional eccentric loops (with a line density), \citet{Pan16} concluded that slower apocentre dust velocities can `pile-up' dust preferentially at apocentre.
The predictions of this 1-D theory appear to match ALMA observations of debris discs such as those of Fomalhaut and HD53143 which both have peak ansae emission towards the direction of their imaged apocentres \cite[see][]{MacGregor17, MacGregor22}, i.e., these exhibit `apocentre glow'.

More recently, \citet{LynchLovell22} modelled three-dimensional debris disc density structures with a forced eccentricity component and eccentricity gradients, and (in the long-wavelength regime, defined as $\lambda \gg \lambda_\star := hc/k_B T_\star \sqrt{2a/R_\star}$) made three important conclusions.
Firstly, in discs with constant eccentricity, orbits are more radially bunched up at pericentre and spread out at apocentre, resulting in millimetre-wavelength pericentre glow.
Secondly, that \textit{only in the unresolved limit} (i.e., disc emission remains observationally separated from the stellar emission; $\Delta a \ll B \ll a$, for discs with widths/extents $\Delta a$ and semi-major axes $a$, imaged with beams with angular extent $B$) are the results of the classical line-density theory recovered, with apocentre-enhancements achieved in poorly resolved, large radii, narrow discs. 
Conversely, at higher-resolution, pericentre brightnesses are enhanced, where observations are more sensitive to the densely packed dust orbits at pericentre.
Finally, it was shown that disc eccentricity \textit{gradients} can further enhance pericentre glow if these \textit{rise} with $a$ or apocentre glow if these \textit{fall} with $a$.

Since the publication of \citet{LynchLovell22} new eccentric disc observations have been reported \citep[e.g., HD53143;][]{MacGregor22}.
By modelling the disc of HD53143 with free and forced eccentricity components, this new work concludes that HD53143 hosts amongst the most eccentric discs ever observed, finding best-fit parameter estimates of $e_p=0.11$ and $e_f=0.21$ respectively (for the free ($e_p$) and forced ($e_f$) eccentricity components). 
This disc shows evidence of an apocentre brightness enhancement, at a level close to the predictions of \citet{Pan16}, despite being observed at a relatively high-resolution with ALMA, i.e., a beam with a resolution of ${\sim}1.0''$. 
There are key differences however between the models of \citet{MacGregor22} and \citet{Pan16} which complicate their direct comparison that we explore in this paper.

\begin{figure*}
    \centering
    \includegraphics[width=\textwidth]{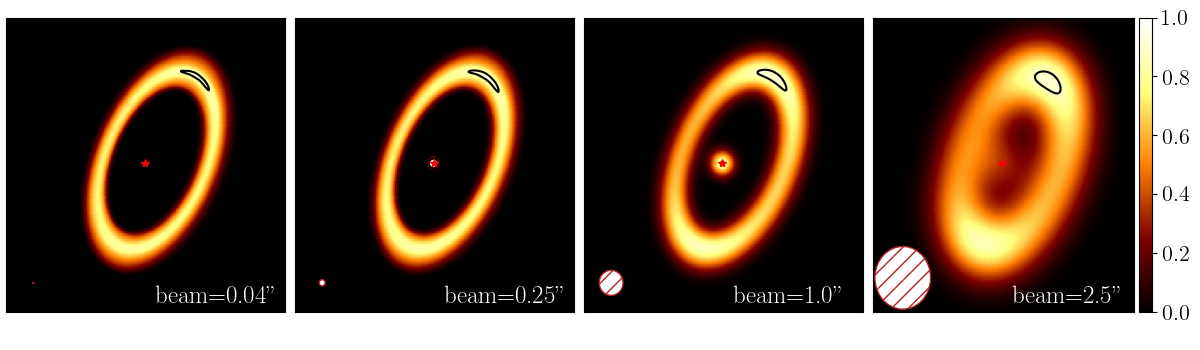} \vspace{-8mm}
    \caption{Parametric modelling results for HD53143 with $e_p{=}0.11$ and $e_f{=}0.21$, in the same coordinate system as M22. From left to right we show progressively lower resolution images, at $0.04''$, $0.25''$, $1.0''$ and $2.5''$. In the lower-left of all images we show the beam size associated with the resolution scale. Note that the images have been linearly scaled to the peak disc flux, and as such our images are unitless. All contours are shown at 95\% of the peak disc brightness.}
    \label{fig:HD53143}
    \vspace{-0.05in}
\end{figure*}

In this work, Paper II of the Eccentric Debris Disc Morphologies series \citep[also see Paper I;][from here on referred to as Paper I]{LynchLovell22}, we likewise model the emission of debris discs with a \textit{3-dimensional} density distribution (imaged on 2D grids). 
Here we present results from new models of dust orbits parametrised with both free \textit{and} forced eccentric components, and explore the impact of disc widths on disc brightness morphologies, for the restricted case of a face-on disc with the same parameters derived for HD53143 \citep[as presented in][from here on referred to as M22]{MacGregor22}.
We demonstrate in these models that apocentre enhancements are strongly resolution and disc width dependent, and suggest that a more natural explanation for discs exhibiting strong apocentre glows that these are instead driven by brightness enhancements due to having falling eccentricity gradients, i.e. $\partial e/ 
 \partial a < 0$.
As in our previous work, we use the terms pericentre glow and apocentre glow to refer to the apse in which the \textit{peak surface brightness} appears, though note that inclinations can alter the precise peak emission location. 
For this reason, we conduct our later analyses with face-on models.
In Section \ref{sec:modeltheory} we present our model setup, in \ref{sec:discussion} we discuss the implications of our models, and present our main conclusions in \ref{sec:conclusions}.

\vspace{-0.2in}
\section{A parametric model of a disc with free and forced eccentricity}
\label{sec:modeltheory}

\subsection{Model setup}
\label{sec:modelparams}
The modelling methodology we adopt here follows the same prescription as our Paper I, with a modified surface density distribution that now includes the free eccentricity.
We developed a single non-dynamical, parametric disc model with a top-hat distribution of dust semi-major axes (with user-defined parameters as discussed below) and imaged this with the $\rm{RADMC-3D}$ \citep{Dullemond12} package.
To account for the free eccentricity component, we superimpose multiple families of orbits (with each family corresponding to our constant eccentricity model from Paper I). Each family includes a contribution from a free eccentricity component, $e_{\rm{p}}$, with a fixed argument of pericentre for the free eccentricity, $\omega_{\rm{p}}$. The resulting dust-density is obtained by summing over families which uniformly sample $\omega_{\rm{p}}$.
We adopt this prescription here for N=16,000 orientations of $\omega_{\rm{p}}$, which we tested to ensure sufficient numerical convergence on our image grids (with sizes of 2000x2000 pixels, and a pixel scale of $0.01''$, or 0.18\,au at the distance to HD53143).
In this section we continue by exploring how this new model setup manifests at millimetre wavelengths.

We define a series of parameters required to model optically thin debris disc dust emission within the $\rm{RADMC-3D}$ framework, which unless otherwise explicitly stated, remain constant throughout, with values consistent with M22.
These include the observational wavelength ($\lambda_{\rm{obs}}$, fixed at 1.3\,mm), the distance ($d$, fixed at 18.3\,pc), the peak emission semi-major axis ($a_0$, fixed at 90.1\,au), the disc width ($\Delta a$, variable between 2\% and 50\% of $a_0$), the vertical aspect ratio ($h$, fixed at 4\%), the forced eccentricity ($e_{\rm{f}}$, fixed at 0.21), the free eccentricity ($e_{\rm{p}}$, fixed at 0.11) the argument of pericentre of the forced eccentricity ($\omega_{\rm{f}}$, fixed at either $112.8^\circ$ for the inclined model, or $90^\circ$ for the face-on model, where $\omega_{\rm{f}}$ is defined as the anticlockwise angle from North), the inclination ($i$, fixed at either $65.6^\circ$, or face-on, i.e., $0^\circ$), the position angle ($\rm{PA}$, from North, anti-clockwise, fixed at $156.4^\circ$), the phase offsets in RA and Dec ($\delta_{\rm{RA}}$ and $\delta_{\rm{Dec}}$, fixed at $0.07''$ and $0.04''$ respectively), the minimum and maximum grain sizes ($D_{\rm{min}}$ and $D_{\rm{max}}$, respectively set by the blowout size of a ${\sim}$solar-type star, and by neglecting emission from larger grains than ALMA wavelengths at values of $0.9\,\mu$m and 1.0\,cm), the dust grain density ($\rho$, fixed at $2.7\,\rm{g\,cm}^{-3}$), and the grain size power-law distribution \citep[$\alpha$, fixed at -3.5, as per][]{Dohnanyi69}. 
We fixed the total dust mass ($M_{\rm{dust}}$) at $0.05\,M_\oplus$. Although this parameter was not modelled by M22, since this only acts as a \textit{total} brightness scaling factor in the optically thin limit, and our investigation only considers fractional brightness ratios, we are free to set this parameter to anything as long as the disc remains optically thin.

The dust temperature in our models is determined by the stellar temperature ($T_\star$) and stellar radius ($R_\star$) which define a template \citep{Kurucz79} stellar spectra (in all instances in this work these were fixed as $5250\,$K and $1.0\,R_\odot$ respectively), and the stellar mass ($M_\star$) was fixed as $0.9\,M_\odot$. 
In all models we scale the flux of the star to $F_{\rm{star}}=50\,\mu\rm{Jy}$ (consistent with the stellar flux modelled by M22), and fix the origin of the image coordinate system at the star's location.
Our models have a vertical Gaussian density distribution, defined by the vertical aspect ratio, $h=H/r$, where $H$ is the absolute vertical height of dust at a radius $r$ in the disc. 
This same parameterisation has been applied previously to model sub-mm ALMA observations \citep[see][]{Marino16, Marino18, Lovell21C}, as well as in Paper I and is consistent with both the expected physical distribution of dust above and below the disc mid-plane, and the work of M22.
Our models are projected on to 2-D grids in $(r,\phi)$ space, for which $r=0.0''$ corresponds to the origin (the location of the star).


\vspace{-0.3in}
\section{Disc Modelling: Results and discussion}
\label{sec:discussion}

\subsection{Model verification and comparison}
\label{model:verified}
To verify our model behaviour as correct, we successfully reproduced the global disc emission structure of HD53143, as imaged and modelled by M22.
In doing so, we demonstrate that our models, now including free eccentricity are correctly reproducing the emission of an inclined disc with both free and forced eccentricities, and can accurately predict the structures associated with more general disc scenarios (i.e., those with different parameters, such as $i$, $\rm{PA}$ and $\Delta a$).
We show in Fig.~\ref{fig:HD53143} our model imaged at $0.04''$, $0.25''$, $1.0''$ and $2.5''$, of which the $0.04''$ resolution corresponds to an imaged `full resolution' model of M22 (within ALMA's resolution capabilities) and to progressively lower resolution data, all of which can be achieved with sufficiently deep ALMA observations.
This correctly reproduces the disc's morphology (i.e., an inclined, 90.1\,au narrow ring with a large central cavity, and star offset from its centre), and importantly for our later investigation, apocentre glow (i.e., a brightness enhancement on the north-east disc ansa) in all four resolutions, and at the same level as HD53143's at $1.0''$ resolution.

Determining the best-fit parameters of our model is beyond the scope of our investigation, since the above has shown that using identical parameters to the $10^7$ particle method of M22, we can accurately reproduce the model of HD53143 with a parametric model, which we verified by producing a model image based on the same parametrisation but instead using $10^7$ particles to sample the dust distribution (i.e, consistent with the approach of MM22).
As noted by \citet{Kennedy20}, particulate models of discs imaged at sub-arcsecond resolution require many millions of particles to minimise computational shot-noise error to far below the errors induced by observations (an issue significantly compounded for ever-higher resolution models, which we discuss in Appendix~\ref{appB:particles}).
In verifying our parametric model, we found that our parametric approach thus has a significant computational advantage versus particulate models at high-resolution, where numerical convergence is achieved much faster due to two factors.
Firstly, due to the increasingly higher number of particles required to reduce model noise to sufficiently low levels (see Appendix~\ref{appB:particles}).
Secondly, that our parametric model derives solutions for complete orbit \textit{families} rather than individual particles, which greatly reduces the number of calculations needed to accurately define the density of dust in a circumstellar disc.

\begin{figure*}
    \centering
    \includegraphics[width=0.975\textwidth]{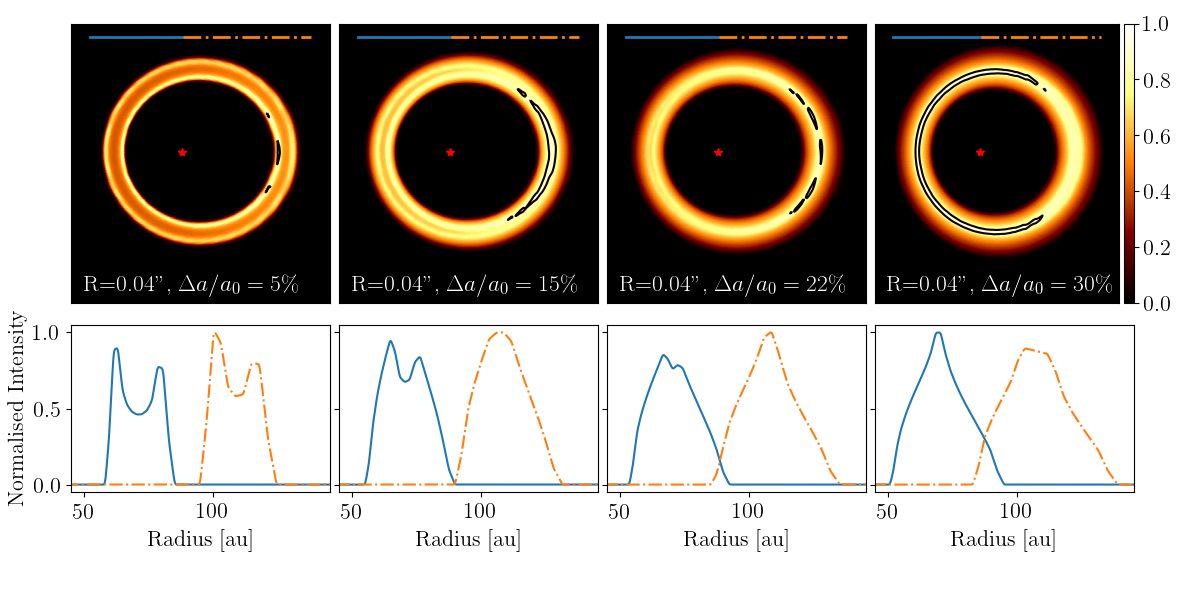} \vspace{-6mm}
    \caption{Top: normalised parametric model images for face-on discs with the same parameters as our HD53143 model i.e., $e_p{=}0.11$ and $e_f{=}0.21$, but with variable ring-widths of 5\%, 15\%, 22\% or 30\%, simulated at $\lambda_{\rm{obs}}{=}1$mm. 
    Contours are set at 95\% of the peak disc brightness.
    Bottom: major axis surface brightness profiles for each image, across both ansae. 
    Initially the maximal brightness of the disc appears on the disc inner and outer edges (apocentre enhanced). 
    As ring width is increased (at fixed $0.04''$ resolution) overlapping dust orbits reduce the dust emission minima, and result in pericentre glow. }
    \label{fig:faceOn_varWidth}
    \vspace{-0.05in}
\end{figure*}

\vspace{-0.15in}
\subsection{Ring width and resolution: the origin of free-eccentricity induced brightness enhancements}
The main aim of our investigation is to determine how disc surface brightness variations can arise at millimetre wavelengths (e.g., such as apocentre and pericentre glows) in systems which have \textit{both} free and forced eccentricities.
For this, we use the example of HD53143 given this is the most eccentric millimetre-resolved disc yet observed and as such makes this disc the most straightforward to visualise how ring widths and free/forced eccentricities interact.
We note that there is otherwise nothing special about our choice of this debris disc.

We start with an identical model to that presented in \S\ref{model:verified} and alter this by i) inclining this to a face-on orientation, ii) rotating the position angle horizontally east-west, and iii) re-producing and imaging the model for a range of narrow fractional widths ($\Delta a/a_0$) from 5\% to 50\%.
This range of widths was selected to ensure that we sampled all plausibly `narrow' discs (typically defined as those having fractional widths below 50\%), and down to the lower limit at 5\%, below which we deem the model behaviour too noisy for reasonable analysis of disk structures.
Observations of debris discs have found there to be a wide range of widths, including broad discs \citep[i.e., with fractional widths likely in excess of 50\%, e.g., q$^1$Eri, and $\kappa$CrB, see][respectively]{Lovell21C, Lovell22}, and discs as narrow as Fomalhaut \citep[i.e., with a modelled fractional width of ${\sim}8\%$, see][]{MacGregor17}.
Here we focus on narrow discs exclusively.

We present in Fig.~\ref{fig:faceOn_varWidth} a subset of these models; all shown at high resolution ($0.04''$) for discs with fractional widths of 5\%, 15\%, 22\% and 30\% (noting that 22\% is the MM22-determined fractional width of HD53143).
This figure demonstrates an important observational point: as the fractional width increases at a fixed (high) resolution, the surface brightness varies through the disc, e.g., with the initial apocentre glow (in the narrow discs) altered to a pericentre glow (in the broader disc) at millimetre wavelengths.
Instead of the millimetre-wavelength apocentre glow being an effect due (solely) to resolution and wavelength, here we show that this is the result of overlapping dust orbits for dust belts modelled with both free and forced eccentricity components.
In the radial profiles shown in Fig.~\ref{fig:faceOn_varWidth}, it can be seen that these images exhibit emission profiles that, when most narrow, are double peaked, consistent with the theoretical models of \citet{Kennedy20} and \citet{Rafikov23} (and from their associated images, that azimuthally, these radial minima are present throughout the disc).
However, due to the different physical widths of discs as measured at their apocentre and pericentre directions (for fixed eccentricities), the gap between the double-peaked apocentre emission reduces before the gap on the pericentre side (e.g., for this model, for 15\% width discs, the apocentre radial peaks now overlap, whereas the pericentre ansa remains double-peaked). 
This has the effect that for optically thin, eccentric discs with sufficiently narrow widths, the surface density of dust (and thus emission brightness) on the apocentre side rises more sharply than the pericentre side.
At a fractional width of 22\% the pericentre ansa remains double-peaked, though with a smaller separation.
However by 30\%, the double-peaked pericentre emission has ceased, and pericentre dust orbits now overlap constructively, producing a sharp increase in the pericentre emission brightness.
On the apocentre however, dust orbits are instead being radially spread out, reducing the apocentre dust surface density.
This has the effect, that (for face-on discs with all of HD53143's other parameters held fixed, and fixed $0.04''$ resolution), fractional width increases above a given threshold width induce \textit{pericentre glows at millimetre wavelengths}.

\begin{figure}
    \centering
    \includegraphics[width=1.0\columnwidth]{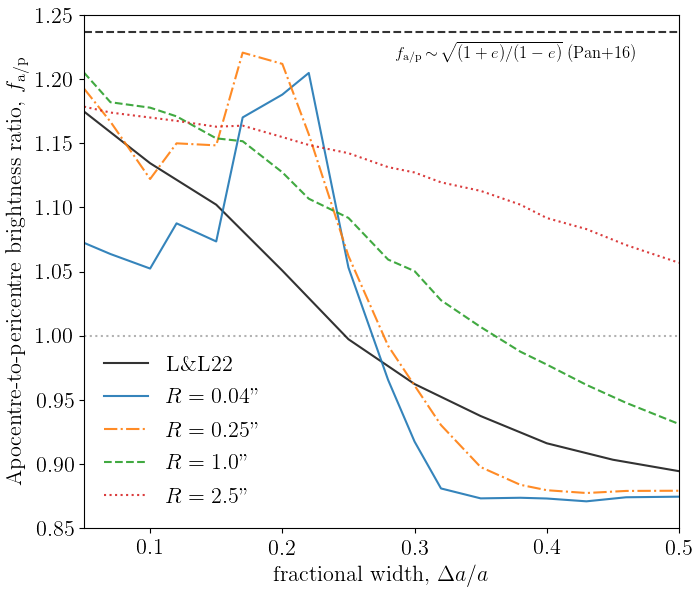} \vspace{-6mm}
    \caption{Peak apocentre-to-pericentre brightness ratios, $f_{\rm{a/p}}$ as a function of fractional disc width, $\Delta a/a$, for our model grid. We also plot the expected brightness enhancement from the models of \citet{Pan16} for a disc with fixed eccentricity of 0.21 (black dashed line), and the predictions from our Paper I model for a disc model with only a constant forced eccentricity of 0.21 (i.e., neglecting free eccentricity) as the black line `L\&L22' ($1''$ resolution).}
    \label{fig:fapVarWidth}
    \vspace{-0.2in}
\end{figure}

We generalise this picture in Fig.~\ref{fig:fapVarWidth} by showing all of our modelling results between fractional disc widths of 5--50\%, and at all four resolutions.
This plot shows the peak-apocentre-to-peak-pericentre brightness ratio, $f_{\rm{a/p}}$, as a function of fractional width, demonstrating where the brightness enhancement regimes operate for a face-on disc with parameters otherwise consistent with HD53143.
Initially, in all cases, the peak brightness of the narrowest discs appear on the apocentre side, although at the start of this scenario, the number of overlapping dust orbits on the apocentre side is low (with a correspondingly smaller brightness enhancement).
As the orbital overlap of eccentric dust enhances more rapidly on the apocentre side than the pericentre side however, this disc side sees a sharp brightness enhancement over the pericentre side, peaking approximately in the region 15--25\% for sub-arcsecond resolutions, before this rapidly falls, and by 30\% may only achieve pericentre brightness enhancement.
We note here that the best-fit M22 value of 22\% sits inside this 15--25\% apocentre-enhancement range. 

We also show on Fig.~\ref{fig:fapVarWidth} the predictions of the brightness ratio with the models of \citet{Pan16} (black-dashed line) and the models of Paper I for a \textit{constant} eccentricity disc with only fixed forced eccentricity (convolved with a $1.0''$ circular Gaussian beam, black-solid line, `L\&L22') for millimetre wavelengths. 
We note that only in the unresolved limit does the \citet{Pan16} predicted $f_{\rm{a/p}}{\sim}\sqrt{{(1+e)/(1-e)}}$ remain valid, under the assumption that $e$ is dominated by the forced (constant) eccentricity component.
On the other hand, we show that the behaviour of our constant eccentricity Paper I model follows broadly the same profile as the $R=1.0''$ free and forced component model, predicting apocentre glow in poorly-resolved narrow discs, and pericentre glow otherwise.
We note two additional points.
Firstly, that in the limit of very low fractional widths, our constant eccentricity `L\&L22' model converges on the \citet{Pan16} result as our disc widths become unresolved.
Secondly, that our model under-predicts the apocentre enhancement expected of models that include free eccentricity (by between 5--10\%, for the shown fractional widths) but as disc widths increase, our L\&L22 model accurately tends to the same value as the models with both free and forced components.
All of our 2D disc model images are provided as Supplementary Information.

The apocentre glow origin in M22's HD53143 debris disc model is the result of overlapping eccentric dust orbits, for which a disc with a narrow fractional width and high free and forced eccentricities induces surface density distributions with a \textit{single} radial peak at apocentre, and \textit{two} radial peaks at pericentre \citep[an interpretation inconsistent with the line-density modelling of][]{Pan16}.
We note therefore that to achieve apocentre glow at a level matching the observations of HD53143 (i.e., approximately $15\%$, MM22) may require significant fine-tuning of the disc width, given the results we present here.
We have shown that to achieve 10--20\% apocentre brightness enhancements at $1.0''$ resolution (consistent with the M22 observations) models are forced into the region of parameter space where eccentric dust orbits overlap to induce a single peak at apocentre and two peaks at pericentre, only possible with disc widths in the range $15\% {<} \Delta a/a {<} 25\%$, for fixed (high) free eccentricity.

This description may not be a true reflection of what is going on physically however, since the disc width could lie outside of this width range. 
In Paper I we showed that observable apocentre dust enhancements are induced by \textit{falling} eccentricity gradients (e.g., consistent with forcing from an eccentric planet internal to the disc).
This may be a more natural explanation for the origin of HD53143's apocentre glow, since such an internal eccentric perturber/s can provide the force necessary to shape the disc's eccentric structure and apocentre glow (over its ${\gtrsim}$Gyr lifetime) without requiring such stringent conditions on the disc width.
As such, higher-resolution ALMA observations that resolve HD53143's disc width could provide sufficient data to model the disc's underlying eccentricity distribution and determine whether or not this is consistent with either i) constant eccentricity parameters, or ii) internal planet-forcing. 

\subsection{Implications for inferred planetary architectures}
Our results demonstrate that based on the width of an eccentric disc with free and forced eccentricity, the non-overlapping orbits of (radially double-peaked) eccentric dust in very narrow discs can result in local radial emission minima over \textit{all} azimuthal angles at high angular resolution (see model $\Delta a/a=5\%$ in Fig.~\ref{fig:faceOn_varWidth}).
This is analogous to the sub-structure observed in protoplanetary discs ascribed to planet-carving \citep[e.g., such as those in the DSHARP survey;][]{Andrews18} and more recently in optically thin debris discs observed with annular emission minima \citep[e.g., such as HD92945, see][]{Marino19}.
Whilst further detailed work should be conducted to understand the implications for debris discs more generally (accounting for different disc radii, widths and eccentricities), our results show that structures that are readily ascribed to embedded planet-carving may be instead consistent with dynamical expectations from discs with free eccentricities (i.e., in the absence of embedded planets).


\vspace{-0.2in}
\section{Conclusions}
\label{sec:conclusions}
In this work we have developed models of HD53143, to demonstrate the effect of free and forced eccentricities on debris disc morphologies. 
This work is a direct extension to, and supports the findings of our previous study (Paper I) of \citet{LynchLovell22} by demonstrating that interpretations of eccentric disc structures based on line density models remain valid only in a very limited set of circumstances, which are increasingly unlikely to be met as debris disc observations become better resolved.
In this study we conclude:
\begin{enumerate}
 \item in general, by including constant free and forced eccentricities in disc models, at millimetre wavelengths discs may exhibit surface brightness variations that result in \textit{either} pericentre or apocentre glows at millimetre wavelengths;
 \item that pericentre/apocentre glows in such models are a result of preferential overlapping eccentric dust orbits (with uniformly distributed arguments of pericentre of the free eccentricity) i.e., not from the pile-up of lower-velocity dust (i.e. apocentre hang-time);
 \item that in the case of HD53143, new observations and detailed modelling is needed to determine the origin of its apocentre glow (which may either be due to a falling eccentricity profile, or a narrow ring with (constant) free and forced eccentricity);
 \item that the structures induced by free and forced eccentricity models may induce observational signatures in debris discs, consistent with those ascribed to planet carving.
\end{enumerate}

\vspace{-0.2in}
\section*{Acknowledgements}
We thank Grant Kennedy for useful discussions on eccentric disc modelling and comments on an earlier paper draft, and the referee for constructive suggestions that improved both the clarity and quality of our paper.
J. B. Lovell thanks the STFC (through a postgraduate studentship) and the Institute of Astronomy, University of Cambridge, for partly funding this work through a Summer Research Assistantship, and the Smithsonian Institute for funding via a Submillimeter Array (SMA) Fellowship.
E. M. Lynch thanks the Science and Technologies Facilities Council (STFC) for funding this work through a STFC studentship, and the European Research Council (ERC). This research was supported by STFC through the grant ST/P000673/1 and the ERC through the CoG project PODCAST No 864965. This project has received funding from the European Union’s Horizon 2020 research and innovation program under the Marie Skłodowska-Curie grant agreement No 823823. 

\vspace{-0.2in}
\section*{Data availability} \label{dataavail}
All model files and surface density structure code will be made available via JBL's github: \url{https://astrojlovell.github.io}.


\vspace{-0.2in}
\bibliographystyle{mnras}
\bibliography{mnras_template.bbl} 


\appendix

\section{Particle Model Error Analysis} \label{appB:particles}
In this Appendix we briefly investigate one complication introduced by modelling high-resolution discs with particle models, by measuring the differences between images produced by models based on a dust surface density distribution and those produced when this surface density distribution is sampled with a finite number of particles.
For illustrative purposes we considered only a single \textit{circular} Gaussian disc model (i.e., $e_p{=}0$ and $e_f{=}0$, with disc parameters all otherwise fixed to those presented in \S\ref{sec:modelparams}) with a 20\% fractional width ($\Delta a /a$), and sampled either $10^4$, $10^5$, $10^6$, $10^7$, $10^8$, or $10^9$ particles (in the particle Gaussian sampling method).
We ran these dust density distributions through the same imaging pipeline described in \S\ref{sec:modelparams}.
All model images were then convolved with either $0.04''$, $0.25''$, $1.0''$ and $2.5''$ Gaussian beams, and for each resolution scale, we subtracted each particle model image from the Gaussian function model image to obtain model residual maps. 
Finally, we then measured the rms within a region defined by the ${\pm}3\sigma$ width of the Gaussian function (i.e., with $\sigma=\Delta a / 2\sqrt{2\ln{2}}$).

\begin{figure}
\includegraphics[width=1.0\columnwidth]{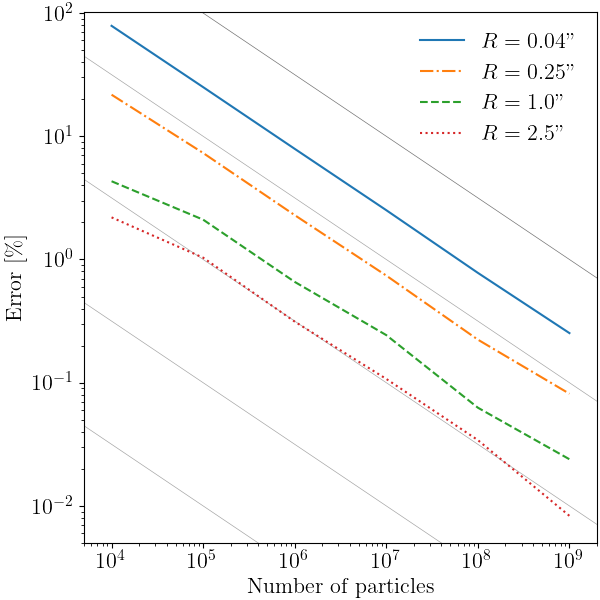}
\caption{Model induced errors as a function of particle number, for the four model resolutions considered within this paper. 
We define the error as the rms of the residual image emission between the two described disc images within an extent ${\pm}3\sigma$ widths from the centre of a circular Gaussian function ($a_0$) for discs with fractional widths of $\Delta a / a = 20\%$. 
These model residual errors are consistent with that due to sampling variance, which falls as RMS$\propto N_{\rm{particles}}^{-1/2}$, which we show as thin grey lines.}
\label{fig:particlesError}
\end{figure}

We show in Fig.~\ref{fig:particlesError} the outcome of the measured residual rms values (i.e., the model error) as a function of particle number. 
Firstly, we demonstrate that for increasing particle number, the rms errors fall following the relationship RMS $\propto 1/\sqrt{N_{\rm{particles}}}$ as expected when the error is dominated by the sample variance.
Therefore for ever higher-resolution observations, the number of particles needed to accurately model debris disks increases. 
This has important observational implications for future measurements and modelling exercises, e.g., for the ALMA Large Program ARKS (Marino et al. 2023, in preparation), or for HD53143. 
ARKS is scheduled to resolve the structure of 18 debris discs on scales ranging from $0.03-0.8''$, and achieve per-beam SNRs in the range of 5--10.
Thus, for particle-based models to accurately model these systems, i.e., producing modelling errors ${\gtrsim}50\times$ smaller than observational errors, these will require rms errors of 0.1--0.2\%. 
Based on Fig.~\ref{fig:particlesError}, for an approximate average ARKS resolution of $0.25''$, this will require models with ${\gtrsim}10^9$ particles.
In the case of HD53143, which was observed with approximately an average per-beam SNR of 7 (across the complete disk extent), resolving this with a factor of 4 resolution improvement (i.e., from the existing $1.0''$ MM22 resolution to $0.25''$), our investigation implies that ${\gtrsim}10^8$ particles would then be needed to ensure that observational errors dominate modelling errors (an order of magnitude larger than used in MM22).

In summary, we have shown that as higher-resolutions are achieved observationally, the computational cost of modelling these with existing particle-based models increases approximately quadratically. This incentivises adopting methods that scale better with increased resolution.

\section{Supplementary Material} \label{appA:modelsuite}
In our Supplementary Material we provide 2D images of the model suite used to interpret the brightness enhancements in \S\ref{sec:discussion}, and in the online version a `.gif' movie of each row presented in this grid. 
All of these models will be made publicly available via \url{https://astrojlovell.github.io}.

\begin{figure}
\includegraphics[width=1.0\columnwidth]{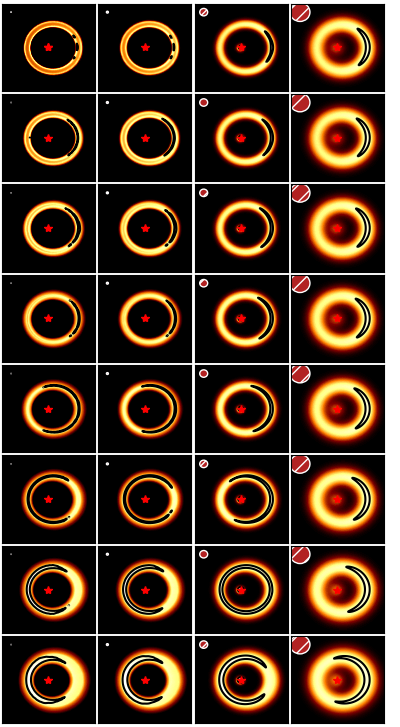}
\caption{Two-dimensional face-on models of debris discs with increasing fractional widths (top to bottom, with fractional widths of 5\%, 10\%, 15\%, 20\%, 25\%, 30\%, 40\%, and 50\%) and decreasing observational resolution (left to right, with resolutions of $0.04''$, $0.25''$, $1.0''$ and $2.5''$). Apocentre glow is seen in all of the low-resolution images, however either pericentre glow \textit{or} apocentre glow is demonstrated at the other three resolutions, depending on their fractional widths and resolution scale. 
In the top-left of each plot we show the beam size (resolution scale).
All contours are shown at 95\% of the peak disc brightness.}
\label{fig:orbits}
\end{figure}

\bsp	
\label{lastpage}
\end{document}